# Fast, Flexible, Polyglot Instrumentation Support for Debuggers and other Tools

Michael L. Van De Vanter[a], Chris Seaton[a], Michael Haupt[b], Christian Humer[a], and Thomas Würthinger[a]

a   Oracle Labs
b   Work done while at Oracle

**Abstract**   *Context:* Software development tools that interact with running programs such as debuggers, profilers, and dynamic analysis frameworks are presumed to demand difficult tradeoffs among implementation complexity (cost), functionality, usability, and performance. Among the many consequences, tools are often delivered late (if ever), have limited functionality, require non-standard configurations, and impose serious performance costs on running programs.

*Inquiry:* Can flexible tool support become a practical, first class, intrinsic requirement for a modern high-performance programming language implementation framework?

*Approach:* We extended the Truffle Language Implementation Framework, which together with the GraalVM execution environment makes possible very high performance language implementations. Truffle's new Instrumentation Framework is language-agnostic and designed to derive high performance from the same technologies as do language implementations. Truffle Instrumentation includes: (1) low overhead capture of execution events by dynamically adding "wrapper" nodes to executing ASTs; (2) extensions to the Language Implementation Framework that allow per-language specialization, primarily for visual display of values and names, among others; and (3) versatile APIs and support services for implementing many kinds of tools without VM modification.

*Knowledge:* It is now possible for a client in a production environment to insert (dynamically, with thread safety) an *instrumentation probe* that captures and reports abstractly specified execution events. A probe in fully optimized code imposes very low overhead until actually used to access (or modify) execution state. Event capture has enabled construction of numerous GraalVM services and tools that work for all implemented languages, either singly or in combination. Instrumentation has also proved valuable for implementing some traditionally tricky language features, as well as some GraalVM services such as placing bounds on resources consumed by running programs.

*Grounding:* Tools for debugging (via multiple clients), profiling, statement counting, dynamic analysis, and others are now part of GraalVM or are in active development. Third parties have also used Truffle Instrumentation for innovative tool implementations.

*Importance:* Experience with Truffle Instrumentation validates the notion that addressing developer tools support as a *forethought* can change expectations about the availability of practical, efficient tools for high-performance languages. Tool development becomes a natural part of language implementation, requiring little additional effort and offering the advantage of early and continuous availability.

# The Art, Science, and Engineering of Programming

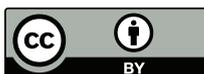

 **Introduction**

A frequent programmer complaint about new programming languages (or new implementations of old languages) is that runtime-based *tools* such as debuggers, profilers, and dynamic analyzers typically arrive late, if ever. When they do arrive it is often with functional limitations, inconvenient configuration requirements, and runtime overhead that constrains their use.

Tools are too often treated as an *afterthought* (e.g. JVMPI, JVMTI for Java [26]), but not always. Gabriel reminds us that some of the earliest and most influential programming languages, for example Lisp (1965) [23], Smalltalk (1980) [13], and Self (1989) [2], were actually programming *systems* that exhibited little distinction between language and tools [12]. The shift in focus toward compilers as separate artifacts (which optimize utilization of *expensive machines*) and away from programming tools (which optimize utilization of *expensive people*) came at a cost, reflected in the complaints recited above and an unacceptable loss of productivity.

We argue that programmers should "have it all" [38]:

1. Tools that require dynamic access to runtime state should be conveniently available at all times;
2. Runtime overhead should be low enough that tools can be activated in most situations, even in deployed systems where performance is important;
3. New languages and language implementations should arrive with tools; and
4. It should be easy to build tools without requiring changes to language implementations, so that new ones can be added as experiments or to adapt to new or local development processes.

Getting there may sound expensive, if not prohibitive, but when addressed with *forethought* there are strategies for managing those costs:

- Hooks that support a wide variety of tools can be built directly into a high-performance execution environment;
- The environment's optimization technologies can be used to minimize runtime overhead introduced by hooks;
- Language-agnostic tool support can be shared, amortizing its cost over many languages and reducing per-language specializations;
- A rich API for client tools, combined with language-agnostic support libraries (e.g. frameworks for debugging and dynamic analysis), amortizes its cost over many tools and simplifies the creation of new ones.

Adopting these strategies for "having it all" would produce benefits beyond the ritual complaints. One would be to make it more attractive to create new Domain Specific Languages, which seldom enjoy tool support. Another would be to pave the way toward uniform tool support across languages, both when used singly but also when combined in increasingly important *polyglot* applications that combine code in multiple languages.

This paper presents the Truffle Instrumentation Framework, an existence proof that these strategies can produce the kinds of outcomes listed above. It does so in the context of the open source GraalVM project at Oracle Labs [27]. GraalVM leverages common infrastructure and advanced optimization and compilation technologies to enable programming language implementations with competitive performance [42]. Truffle Instrumentation extends GraalVM's Truffle Language Implementation Framework with implementations of the strategies cited above.

Tool support is built into GraalVM and is always available.

The critical runtime cost of tool support in fully optimized code, the capture and reporting of execution events at a specified program location, is near zero[1] when not active, is measurably very low when active, and returns to near zero after deactivation.

All Truffle languages[2] implemented at Oracle Labs now support required per-language specializations at very little additional development cost.

The list of tools implemented using Truffle Instrumentation is large and growing. A Debugging API bundled with the platform operates uniformly across all Truffle languages, either singly or when combined using GraalVM language interoperation. It can be used from the NetBeans IDE [28], Chrome DevTools [15], GraalVM's experimental REPL-style shell, as well as other experimental systems. Functionality includes standard stepping, conditional breakpoints, stack traversal, evaluation in context, and more. Debugging also serves as motivation and running example.

Truffle Instrumentation also supports dynamic measurement such as profiling and code coverage. The NodeProf framework for dynamic analysis uses Truffle Instrumentation to create an alternate implementation of Jalangi [35] that performs up to three orders of magnitude better than the original [37]. Some language features that are traditionally difficult to implement (for example Ruby's set_trace_func [9]) are straightforward to support with better results than other platforms. Finally, third parties are using Truffle Instrumentation for implementing innovative tools.

Section 2 of the paper begins with background on GraalVM's essential features, followed in Section 3 by a summary of key insights and design ideas. Section 4 describes how probes are implemented, and argues that the critical performance overhead is reasonable. Section 5 summarizes the small number per-language specializations required from language implementors, and Section 6 describes flexible APIs for tool implementors. Section 7 summarizes results to date with the Truffle Instrumentation, and Section 8 reviews significant related work. Section 9 presents the current and future outlook for the project, followed by conclusions in Section 10.

## 2 Background: GraalVM

Graal is an advanced compiler from Java bytecode to machine code, written in Java, originally developed in the context of the Maxine Research VM [39], that among other advantages exposes a Java API for controlling compilation. Truffle is a Java framework for language implementations expressed as self-optimizing Abstract Syntax Tree (AST) interpreters. Graal recognizes Truffle usage and aggressively optimizes a Truffle-based interpreter, using partial evaluation, speculation and inlining compilation to generate efficient machine code [43]. These two technologies are the core of GraalVM, a productive *platform* for creating very high performance programming language implementations [27].

---

[1] Appendix B will make clear what is meant here by "near zero".

[2] Language implementations based on GraalVM's Truffle Language Implementation Framework are referred to as *Truffle languages*.

*Guest language* implementation using GraalVM is greatly simplified by reusable host services: dynamic compilation, automatic memory management, threads, synchronization primitives, a memory model, and opportunities for low overhead cross-language interoperation. Truffle languages currently include JavaScript, R [36], Ruby [33], and LLVM-based languages [30], among others. Figure 1 summarizes the structure of a deployed Truffle language application.

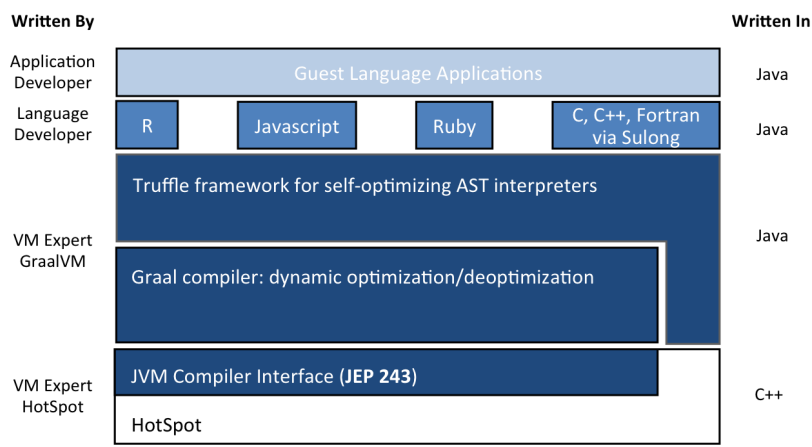

■ **Figure 1** Executing a guest language application with GraalVM technology.

A Truffle language implementation begins with an *AST interpreter*, a straightforward technique that by itself performs poorly. Truffle dynamically optimizes specific nodes by thread-safe replacement with specialized versions.[3] For example, the newly created Truffle AST in Figure 2 is populated by uninitialized ("U") nodes. During execution, nodes are replaced by type-specific nodes, such as the ones exclusively for Integer ("I"). Each specialization is accompanied by a *guard* that verifies on each execution the validity of the specialization. A guard failure replaces a specialized node to a more generic version ("G") that handles all possible cases. No more than a few replacements are permitted at each node, ending with a fully generic implementation if dynamically needed, as suggested by the permissible "Node Transitions" inset in Figure 2.

When an AST stabilizes, GraalVM *dynamically optimizes* [5] its interpreter, using *partial evaluation* [11, 41] to produce highly specialized machine code. Guard failure in optimized machine code triggers dynamic *deoptimization* [18], transferring execution back to the interpreter without loss of execution state (Figure 3). Dynamic deopti-mization frees the compiler to apply speculative optimizations [8] more aggressively, while avoiding compilation overhead for not-yet-seen or slow-path cases.

---

[3] Truffle nodes hold no semantic execution state.

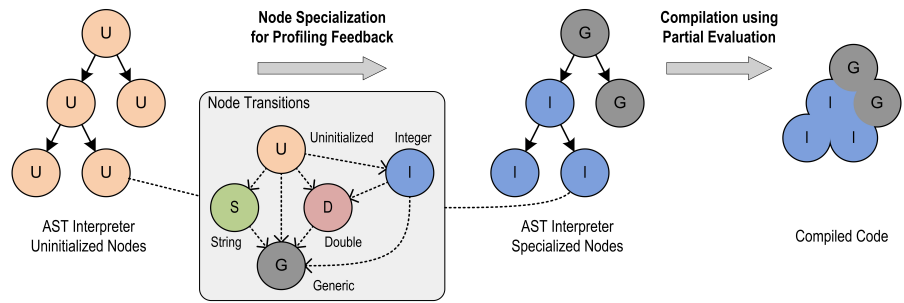

**Figure 2** Truffle ASTs speculate and optimize ...

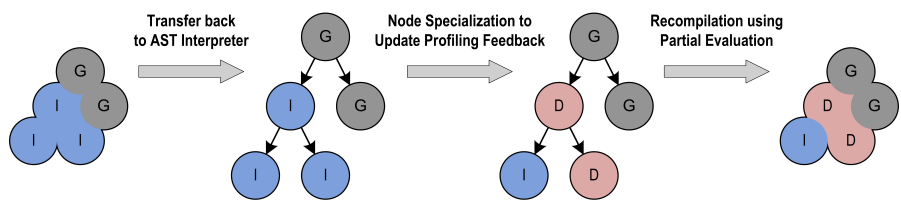

**Figure 3** ... and transfer to interpreter and reoptimize.

## 3 Truffle Instrumentation design overview

GraalVM is a modern, highly productive, all Java, platform for implementing programming languages with very high runtime performance. The Truffle Instrumentation Framework is a GraalVM extension that significantly reduces the cost of building tools requiring dynamic access to Truffle language execution events. Its design addresses the specific concerns mentioned in the introduction: always available, minimal disruption to running programs, minimal additional development requirements for each language implementation, and minimal cost to build a variety of tools.

Two key ideas, inspired by earlier work, informed the architecture of Truffle Instrumentation: capturing events by *interposing* instrumentation code into the flow of program execution (with flexible client access to execution state), and implementing interposition by directly *rewriting ASTs* under interpretation.

A flexible, abstract model for VM-based interposition [17][4] was demonstrated to be sufficient for implementing many different schemes for Aspect Oriented Programming [19], a family of metaprogramming systems that have been performance-limited in practice by lack of support for efficient manipulation of runtime state. Core GraalVM mechanisms make this approach tractable. Well behaved instrumentation code will be fully optimized but can trigger deoptimization when needed, making execution state accessible to instrumentation clients via Java APIs, for example guest language stacks, frames, and frame slots.

---

[4] The only implementation of this model was an experiment that demonstrated its general applicability to various AOP models without concern for performance [32].

The idea of rewriting running code for tooling dates back to Smalltalk [13] breakpoints set by inserting halt statements. Self [2] disrupted traditional debugging by introducing dynamic optimization, but restored it by creating the ability to *deoptimize* dynamically without loss of execution state [18]. Core GraalVM mechanisms make this especially convenient and generalizable. Safe AST rewriting of executing methods, designed to support specialization, can also be used to insert instrumentation nodes. Events at an AST node can be directly associated with the programming language elements represented by the node, for example a statement, call, or expression.

The following three sections describe the aspects of Truffle Instrumentation's design that address requirements for minimization of costs: runtime overhead, additional burden on language implementors, and effort to build tools for every Truffle language.

- The one critical runtime cost is the added time needed to capture an event in fully optimized code and to report it to any interested client. This requires interposing directly into the execution engine, as described in Section 4. An evaluation described in Appendix B shows that this cost is extremely low. Moreover, instrumentation in fully optimized code has near zero [5] overhead until actually used, and it can be removed dynamically, after which overhead eventually returns to near zero.

- The per-language cost for instrumentation is the extra coding needed from language implementors. Section 5 describes specializations needed to support the mostly language-agnostic instrumentation support. A summary shows that the burden is quite modest when compared with the cost of implementing conventional tools for a language. Moreover, the advantage of access to running tools during language development offsets that cost somewhat.

- The cost of developing tools needing access to runtime state can be dominated by the effort it takes to capture events at all, for example by VM modification or bytecode rewriting, in addition to extracting aspects of execution state needed for different purposes. Section 6 describes the versatile Java APIs available to tool builders, where all this can be done at a very high level, in terms of program elements (not implementation artifacts), and in a mostly language- and tool-agnostic fashion.

Figure 4 shows a revised version of Figure 1, annotated to identify additions to GraalVM that implement these ideas.

---

[5] Appendix B will make clear what is meant here by "near zero".

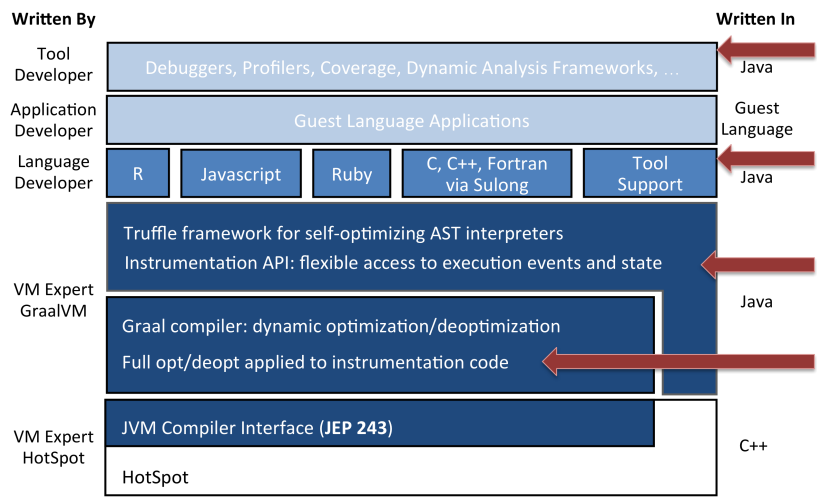

**Figure 4** GraalVM extensions for instrumentation and tools (marked by arrows).

## 4 Low overhead capture of execution events

This section describes how the Truffle Instrumentation Framework supports the dynamic capture and reporting of execution events on behalf of client tools. The implementation is completely language-agnostic.

The paramount goal is to minimize runtime overhead on fully optimized code, achieved via the following three strategies. First, the framework does as little as possible, so instrumented code runs as if a client's event handling code were simply part of the program. The second is to capture events at the granularity of Truffle nodes using fully optimizable extra nodes. Finally, the allocation of extra nodes, and the deoptimization implicitly triggered when they are inserted, is as lazy as possible.

The implementation is presented in a bottom-up fashion, treating in turn:

- Enabling instrumentation by dynamically inserting instrumentation *probes*.
- Routing events from a probe to an interested client by creating a *subscription*.
- Specifying program locations for a subscription by creating a *query*.
- Patching a Truffle language program at a probed location by *injecting* a fragment of guest language code.

The critical performance cost noted earlier, i.e. the overhead added to capture an execution event and report it to clients, is assessed by two experiments reported in Appendix B. Measurements using the instrumentation-based implementation of Ruby settrace [33] show that instrumentation in fully optimized code:

- can be enabled without impact on peak performance when not actually being used;
- has very low impact on peak performance when in use (excluding clients' event handling, of course); and
- reverts to no impact on peak performance when no longer in use.

A more general validation comes from experience with an instrumentation-based dynamic analysis framework for Node.js. NodeProf can be installed without overhead until activated, causes a temporary dip in performance when activated (for re-optimization), and reverts to no overhead when deactivated [37]. Section 7 discusses NodeProf in more detail.

### 4.1 Probes

*Instrumentable nodes* in each AST hold language-provided metadata mentioned in Section 4.3 and described in more detail in Section 5. Truffle *instruments* a node by inserting two additional nodes, as shown in Figure 5.

- An automatically generated *wrapper* ("W") proxies for its child and reports events.
- A language-agnostic *probe node* ("P") dispatches event reports to clients.

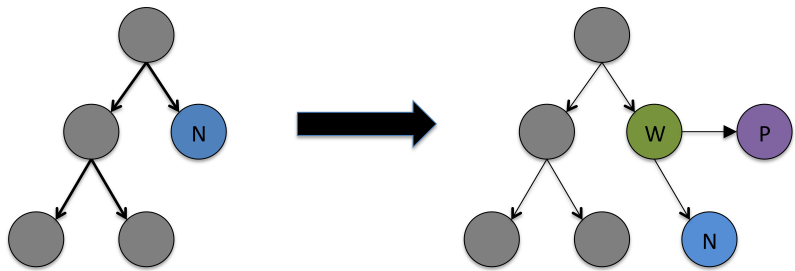

■ **Figure 5**  Probing an AST node with a *wrapper* (proxy) and *probe node* (dispatcher).

The wrapper reports an event to the probe node just before the wrapped child executes and reports another event just after the child executes. Listing 1 summarizes those event signatures, discussed in more detail in Section 6.

■ **Listing 1**  Event signature

```
1    void onEnter(EventContext, Frame)
2    void onReturnValue(EventContext, Frame, Object)
3    void onReturnExceptional(EventContext, Frame, Throwable)
```

Truffle AST's thread safety ensures that probing and un-probing can be dynamic and lazy in order to minimize memory footprint.

Node replacement implicitly triggers deoptimization for any dependent compilations [3, 41], which will eventually be reoptimized. Optimization eliminates unused propagation code, such as the newly inserted probe in Figure 5, since it has no clients.

### 4.2 Subscriptions

Truffle Instrumentation connects an interested client ("C") to a probed node by inserting an additional node that manages the subscription ("S"). Figure 6 shows a "chain" of three subscription nodes, one for each of three independent non-interfering clients. A subscription node propagates each event first to its client with an ordinary Java method call, then to its child (successor in the chain) before returning.

As with wrappers and probes, subscription nodes can be inserted and removed safely and are eliminated by optimization when unused.

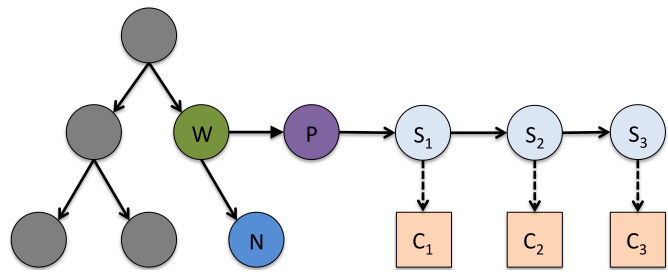

■ **Figure 6** A probed AST node with three chained subscriptions.

### 4.3 Queries

A subscription applies to a *set* of program locations specified by a *query*. Queries combine criteria such as particular sources, kinds of sources, line/column numbers, and syntactic elements, for example "line 42 in mysource.js". A stepping debugger or a coverage tool might use the query "every statement". Section 6 describes the API for queries. Language implementations "mark up" AST nodes with metadata that among other things supports a precise two-way mapping between nodes and source code extents for every syntactic construct (more detail in Section 5.1).

A new subscription (represented by a *binding*) dynamically maintains its set of matching nodes. The implementation challenge is to maintain subscription nodes, as shown in Figure 6, in the presence of changes both to the set of bindings and to the forest of ASTs. A naïve implementation would walk every AST when the set of bindings changes, and would review every binding when a new AST appears.

*Static* optimizations rely on the fact that the source code specified for a typical AST root node subsumes the source for all children.[6] One root check often shows that a query could never match any nodes in an AST.

*Dynamic* optimizations rely on the probability that probed nodes are unlikely to execute before another change to the binding set, if ever. GraalVM optimization uses Truffle `Assumption` objects to track dependencies between method compilations and "assumptions" that the complier has exploited (including assumptions in inlined methods) [3]. Each AST root node holds such an object that (until it is explicitly *invalidated*) represents the fact that the AST has not been modified for the lifetime of the `Assumption`. Any instrumentation change that affects an AST marks the affected probe for lazy modification and invalidates the root node `Assumption`. GraalVM then marks every affected method compilation for lazy replacement and attends (via deoptimization) to cases where an invalidated method happens to be on the stack.

The pseudo-code listing in Appendix A summarizes subscription maintenance.

---

[6] Non-nested sources occur in some languages, handled by a special mark on the AST root.

### 4.4 Code injection

The subscriptions described in Section 4.2 notify clients via ordinary Java method calls. Clients are given access to execution state via Java APIs described in Section 6.

A second form of subscription lets clients *inject* a fragment of Truffle language code[7] into an AST (Figure 7). This kind of subscription responds to an event by executing the fragment in the context of the probed node. Clients have the option to capture return values or exceptions, but neither are propagated into the surrounding program. GraalVM optimizes the fragment as an undistinguished part of the AST. Important use cases for code injection include optimizable breakpoint conditions, tracing, and dynamic performance analysis.

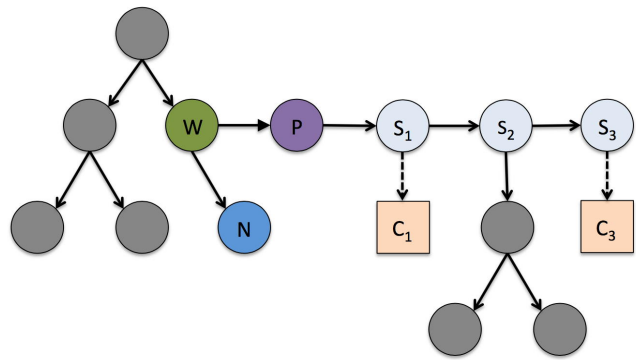

■ **Figure 7**   A probed node with injected code.

## 5   Per-language support requirements

Truffle Instrumentation lowers the per-language implementation cost of tools through sharing of language-agnostic framework code. Language implementors need provide only language-specific *specializations* by implementing a small number of *service provider* APIs, described here informally but completely in the following categories:

- 5.1 *AST Markup*: per-node metadata (source, syntax) meaningful to programmers.
- 5.2 *Visibility*: which implementation elements (e.g. frames, slots) are meaningful to programmers.
- 5.3 *Presentation*: how to display implementation elements (e.g. names, values).
- 5.4 *Expression Inputs*: node value dependencies on other node's return values.
- 5.5 *Patch*: ability to inject an AST fragment into a running method.
- 5.6 *Exceptions*: language diagnostics wrapped in language-agnostic exceptions.

Language implementors who provide specializations early will enjoy the benefit of working tools throughout development.

---

[7] It is a current, but not necessarily permanent, requirement that the AST and injected fragment belong to the same language.

### 5.1 AST Markup

GraalVM treats all AST nodes, including those supporting instrumentation, very much the same. This is an enormous advantage for execution and optimization, but not for people. Programmers think of code interchangeably as either *source text* or language-specific *program elements* (e.g. statements, expressions, blocks), depending on the mental task at hand, but never as ASTs [4]. Truffle language implementors support human tool users by "marking up" language-specific AST nodes with metadata needed for query matching (Section 4.3) and other purposes in two ways: mapping nodes to source text and mapping nodes to syntactic program elements.

**Source Attribution** Every Truffle AST node that represents a syntactic program element provides source information by overriding `Node.getSourceSection`. A `SourceSection` describes the node's corresponding text location precisely (first and last character position) and makes source text available if needed. It also supports two-way mapping between AST and text positions, needed by debuggers for setting breakpoints and visualizing halted locations. Truffle *source attribution* is more thorough than the customary statement line numbers, so for example a debugger could easily be configured to step through expression evaluation with precise visual feedback.

**Syntactic Tags** A debugger must know at which syntactic elements (AST nodes) users expect execution might halt. Language implementors associate a symbolic *tag* with those nodes by overriding `Node.isTaggedWith(Class<?> tag)` to return *true* for `StatementTag`. Other tags currently in use include `CallTag`, `ExpressionTag`, and `RootTag`. More are being added as the suite of built-in services grows.

### 5.2 Visibility

Clients of the Debugging API can enumerate stack frames and frame slots using Truffle's language-agnostic abstractions `Scope`, `TruffleLanguage.findTopScopes(Context)`, and `TruffleLanguage.findLocalScopes(Context, Node , Frame)`. Language implementations, however, sometimes use frames and slots for purposes that correspond to nothing useful or intelligible to a programmer. Truffle Instrumentation introduces the notion of *internal* elements and asks language implementors to identify instances that would be counterproductive to reveal under normal circumstances, for example:

- *Sources* that contain no code sensible to a programmer, for example implementations of language *builtins*.
- *Frames* that correspond to no visible program call, for example Ruby control constructs implemented as method calls.
- *Frame slots* and *object members* that hold implementation-related state, for example intermediate values.

The *internal* designation is independent of underlying implementation and easily changed for improved usability. It reflects a *judgement* that programmers would not benefit (and might be mystified) seeing such an element during ordinary interaction.

Instrumentation clients are free to ignore the designation, however, for example tools specialized for language implementors.

## 5.3 Presentation

Debuggers and other tools often display program information beyond simple source text, for example runtime objects and values. How these should appear is language-specific and sometimes a matter of *convention* based on what programmers expect.

As with the *internal* designation, language-specific presentation is independent of other considerations and easily modified to increase usability. Examples of language-specific presentation that might appear in a debugging session include:

- *Node associated names* such as Method/Procedure name on a root, and Language name on any node, via a key/value property protocol. These are important for stack displays, especially in polyglot debugging sessions.
- *Variable names* via the `Scope` interface using a key/value property protocol.
- *Values*, which are known only as instances of `Object` in GraalVM services, via `TruffleLanguage.toString(Context, Object)`. Important examples include numbers, especially floating point, and distinguished values such as each language's null.
- *Objects*, which can be examined using Truffle's *foreign object* support for *language interoperation* [16]. A key/value property protocol makes available member names and values, and properties such as as writability.

## 5.4 Expression Inputs

A family of important dynamic program analyses depend on tracking *data dependency*: identifying specific values used when computing other values. Examples include dynamic analysis frameworks such as Jalangi [35] for detecting performance bottlenecks in Node.js programs, and dependency analysis for security taint tracking. Truffle language implementors support these by annotating nodes that compute values (*expression nodes*) with information about inputs used in the computation (e.g. child nodes in simple expressions). The NodeProf analysis framework, mentioned in Section 7.2, depends critically on this information.

## 5.5 Patch

Two common debugger interactions are to evaluate a new code fragment, either one time only in the execution context where a program is halted or by injecting it into an AST as described in Section 4.4, for example as a breakpoint condition. Both depend on implementation of method `TruffleLanguage.parseInline(Source, Node, MaterializedFrame)`. This method translates a string into an AST fragment in the static linguistic and lexical context of a specified node and in the dynamic context of some frame on the current execution stack. In both scenarios Truffle Instrumentation patches result into the AST. In the first scenario the fragment is evaluated and immediately removed; in the other it remains on behalf of a client subscription.

**5.6 Language exceptions**

Language implementations are obliged to accompany exceptional returns with instances of the interface `TruffleException`. Each instance contains the information programmers wish to see, for example source location, diagnostic text, the language reporting the error, and access to the stack. They also include internal diagnostics, for example whether an error is syntactic, runtime, or internal to the language implementation. For non-error returns, the exception describes how the program terminated and whether it is an explicit program request to exit.

## 6  Versatile client API

Gaining dynamic access to execution state often requires understanding a great deal about VM internals and sometimes tool-specific VM modifications. Truffle Instrumentation is designed to eliminate those requirements for conventional tools and to encourage the design and experimentation with new kinds of tools. The client API, all in Java, offers tools flexible, language-agnostic, access to execution events and state.

This section demonstrates those attributes with a simple running example, using core client abstractions, derived from the GraalVM Debugging API implementation. Another validation of the client API is the diverse and growing set of clients implemented using Truffle Instrumentation, described in Section 7.2.

**Instruments**    An instrumentation *client* extends class `Instrument` and is installed by placement on the class path of a GraalVM execution environment. Listing 2 contains the skeleton for a simple instrumentation-supported debugging service. The framework activates `Instruments` on demand (method `onCreate()`), when requested by a client, and deactivated (method `onDispose()`) when no clients remain. A newly activated instrument receives the argument `Env` for access to basic features of the current execution environment, for example i/o streams and instrumentation support.

■ **Listing 2**    Skeleton debugging service

```
1  @Registration(id = "debug-service")
2  public final class DebugService extends Instrument {
3      @Override
4      protected void onCreate(Env env) {
5          // Set up debugging support
6      }
7      @Override
8      protected void onDispose(Env env) {
9          // Release resources
10     }
11     public void setBreakpoint(SourceSectionFilter,
12                                 ExecutionEventListener) {
13         ...
14     }
15     ...
16 }
```

Listing 3 sketches a simple scenario for a client tool. It starts by requesting access to the debugging service from the GraalVM execution environment (lines 1-3). It then sets up access to some Truffle language code by creating a `Source` instance, using a builder that supports many other options (lines 4-5). Lines 6-9, which set a breakpoint and execute the code, will be described in more detail one line at a time.

■ **Listing 3** Debugging scenario

```
1   PolyglotEngine engine =...; // A GraalVM execution environment
2   DebugService debugService =
3        (DebugService) engine.getInstruments().get("debug-service");
4   Source mySource = // Access to program source
5        Source.newBuilder("myProgram.xx").build();
6   SourceSectionFilter location = ...;// Specify location in mySource
7   ExecutionEventListener callback =...;// Called when program halts
8   debugService.setBreakpoint(location, callback);
9   engine.eval(mySource); // Run program, expect a callback
```

**Filters**    Clients request events in general by describing a set of program locations with an instance of `SourceSectionFilter`. This class is the client implementation of instrumentation *queries*, described in Section 4.3. Listing 4 expands line 6 from Listing 3, where the client describes the location for a breakpoint ("the statement at line 42 in `myProgram.xx`") using some of the options available in a builder.

■ **Listing 4** Specify breakpoint location, line 6 from Listing 3

```
1   6. SourceSectionFilter location = // Specify location in mySource
2          SourceSectionFilter.newBuilder()
3             .sourceIs(mySource).
4             .lineIs(42)
5             .tagIs(StandardTags.StatementTag)
6             .build();
```

All builder specifications must be satisfied for the filter to match a program location. In addition to tags (any number can be specified), filters can currently be specified by (any number of) specific source names, by (any number of) languages, by line ranges, by character ranges, and whether a source has been marked as "internal".

A filter can match any number of locations. For example a coverage tool might request an event at every statement execution by creating a filter that only specifies that tag. The builder pattern ensures that additional kinds of specifications can be supported in the future without breaking API compatibility.

**Event Listeners**    The Instrumentation framework delivers event notifications from a probed program location via calls to instances of `ExecutionEventListener`, using the event signatures mentioned in Section 4.1. Listing 5 expands line 7 from Listing 3, where the client creates a callback listener that captures execution state just *before* the specified statement is executed. Program execution continues when the callback method returns.

**■ Listing 5**  Create breakpoint callback, line 7 from Listing 3

```
1  7. ExecutionEventListener callback = // Called when program halts
2         new ExecutionEventListener() {
3
4         public void onEnter(EventContext context, Frame frame) {
5             // Handle breakpoint halt
6         }
7         public void onReturnValue(EventContext context,
8                          Frame frame, Object result) {}
9         public void onReturnExceptional(EventContext context,
10                          Frame frame, Throwable exception) {}
11     };
```

**Execution state**   Notifications include information about the event's *context*. The EventContext argument appearing in Listing 5 provides static context: AST, specific Node, language, location, text, and associated tags. The Frame provides dynamic context for the current thread, including method locals.

**Subscription**   Returning now to the debugging service implementation first sketched in Listing 2, the revision in Listing 6 shows how it creates a breakpoint in response to the call on line 8 of the scenario in Listing 3. The debugging service calls the instrumentation framework (attachListener()) to create a *subscription*, described in Section 4.2. That subscription will notify the client callback whenever execution reaches the specified program location, which may happen during execution of the code taking place via the call on line 9 of the scenario in Listing 3.

**■ Listing 6**  Debugging service sets a breakpoint, expansion of Listing 2

```
1  @Registration(id = "debug-service")
2  public final class DebugService extends Instrument {
3      Instrumenter instrumenter;
4      EventBinding binding;
5
6      @Override
7      protected void onCreate(Env env) {
8          instrumenter = env.getInstrumenter();
9      }
10
11     public void setBreakpoint(SourceSectionFilter location,
12                               ExecutionEventListener callback) {
13         binding = instrumenter.attachListener(location, callback);
14     }
15     ...
16 }
```

The resulting EventBinding instance is a handle that can be used to cancel the subscription, which otherwise remains active for the lifetime of the instrument. Any number of subscriptions can be created (and disposed) at any time. A filter is immutable and can be shared by many subscriptions. A listener can participate in many subscriptions.

**Instrumentation errors**    Exceptional return from any ExecutionEventListener method is treated as a client implementation failure. The framework captures exceptions, reports them out-of-band, and allows Truffle language execution to continue.

**Code injection**    Breakpoints created in the above scenario are unconditional. Adding conditions depends on instrumentation support for code injection, described in Section 4.4. The language implementation converts a Truffle language textual expression to an AST fragment for attachment to the AST where it will be evaluated each time execution arrives. When fully optimized, the performance effect is equivalent to the source code having been rewritten at each injected location and eventually optimized again. The following paragraphs demonstrate how the debugging service does this.

An instrumentation subscription for code injection requires a *factory* to produce the AST fragments, an implementation of the interface in Listing 7.

■ **Listing 7**   A factory for AST fragment injection

```
1  interface ExecutionEventNodeFactory {
2      ExecutionEventNode create(final EventContext context);
3  }
```

GraalVM invokes the factory lazily, the first time execution reaches each specified program location. The resulting AST fragment, whose root extends ExecutionEventNode, is then patched into the AST as shown in Figure 7, where it will be executed each time program execution reaches the probed node. The factory is invoked lazily so the static context of each location is accessible, for example lexical binding of variables.

The framework ignores return values from injected code (Section 4.4). The debugging service returns clients an instance of the node class shown in Listing 8.

■ **Listing 8**   Conditional breakpoint node

```
1  class CondBreakNode extends ExecutionEventNode {
2      @Child Node conditionNode;
3
4      CondBreakNode(Node node) {
5          conditionNode = node;
6      }
7
8      @Override
9      public void onEnter(Frame frame) {
10         if ((Boolean) condition.call(frame)) {
11             // Handle conditional program halt
12         }
13     }
14 }
```

The field conditionNode holds the AST fragment that evaluates the conditional expression, which the debugging service creates by delegation to the appropriate language implementation. The @Child annotation is one of the Truffle conventions that enables aggressive AST optimizations. For brevity, this listing ignores handling of exceptional or non-boolean returns.

Finally, Listing 9 shows a revised implementation of the example debugging service, this time with a method for creating conditional breakpoints.

■ **Listing 9**   Debugging service with conditional breakpoints, expansion of Listing 2

```
1  @Registration(id = "debug-service")
2  public final class DebugService extends Instrument {
3      Instrumenter instrumenter;
4      EventBinding binding;
5
6      @Override
7      protected void onCreate(Env env) {
8          instrumenter = env.getInstrumenter();
9      }
10
11     public void setBreakpoint(SourceSectionFilter location,
12                                       final String condition) {
13         ExecutionEventNodeFactory factory =
14             new ExecutionEventNodeFactory() {
15                 public ExecutionEventNode create(EventContext context) {
16                     Node condNode =...; // parse condition in context
17                     return new CondBreakNode(condNode);
18                 }
19             };
20         binding = instrumenter.attachFactory(location, factory);
21     }
22     ...
23 }
```

To recapitulate, when a debugging client requests a conditional breakpoint:

- The debugging service creates an ExecutionEventNodeFactory that holds the text of the condition;
- Instrumentation creates a new subscription via a call to attachFactory().
- The factory, when called, first creates an AST to evaluate the condition in context (by delegation to the language implementation) and then returns a new CondBreakNode that wraps that AST.
- Truffle Instrumentation attaches the new CondBreakNode, where it is executed immediately and on every subsequent execution.
- Each time the CondBreakNode is executed, it evaluates the condition and notifies the debugger if true.

**Source events**    Tools also depend on other kinds of events. For example, the listener in Listing 10 creates a subscription that notifies each time a Source is newly loaded into the runtime. That subscription can also be filtered, for example by MIME types.

■ **Listing 10**   A Listener for Source events

```
1  public interface LoadSourceListener {
2      void onLoad(LoadSourceEvent event);
3  }
```

It is also possible to *query* for Source instances that have already been loaded, optionally subject to a filter.

## 7 Initial Results

This section summarizes experience to date with Truffle Instrumentation Framework from two perspectives: inward-facing, with respect to originally stated criteria, and outward-facing, looking at the breadth and diversity of instrumentation-based client tools that have actually been built.

### 7.1 Criteria

The Introduction proposed four criteria around runtime-based tools that would amount to programmers "having it all", and introduced Truffle Instrumentation as an existence proof that it is possible and practical. Previous sections in the paper have argued, each in a different way, that these criteria have been met.

1. *Tools that require dynamic access to runtime state should be conveniently available at all times*. Instrumentation is intrinsic to every Truffle language implementation. Tools can be invoked via command line options, for example some kinds of profiling, or by standard protocols, for example the v8 suite for debugging and profiling. Tools can be added to the GraalVM classpath and discovered dynamically. Multiple tools can be active and not interfere with one another at the instrumentation level.

2. *Runtime overhead should be low enough that tools can be activated in most situations, even in deployed systems where performance is important*. Section 4 argued that the one critical source of runtime is the cost of capturing an execution event and reporting it to clients when running fully optimized. The framework is designed to exploit platform optimizations such as partial evaluation and inlining so that the effect is as if the client code were part of the program being instrumented, which is what's meant by "near zero" overhead. The experiments reported in Appendix B confirm this with specific measurements, and experience with NodeProf, described further in Section 7.2, suggests this holds true in a complex tool.

3. *New languages and language implementations should arrive with tools*. Section 5 informally describes the six categories of per-language specialization required of language implementors for GraalVM's tools to work. None of these requirements require any new knowledge about tools or the underlying execution engine and are evidently orders of magnitude less work than implementing a complex tool from scratch. Moreover, the prospect of access to working tools *during* language implementation acts as an incentive.

4. *It should be easy to build tools without requiring changes to language implementations, so that new ones can be added as experiments or to adapt to new or local development processes*. Section 6 demonstrates by example that traditionally expensive tool behavior can be implemented using a highly versatile API that abstracts over details of the underlying execution engine and most aspects of language implementations.

## 7.2 Clients

A diverse and growing collection of instrumentation-supported tools for Truffle languages has been developed, some within Oracle Labs, some as academic collaboratorations, and some independently. Many work with every Truffle language, either separately or in polyglot programs using *language interoperation* [16].

### 7.2.1 Debugging
The Debugging API is a core GraalVM service, built using low-level support provided by Truffle Instrumentation, augmented by a library of language-agnostic debugging-related services. For example, breakpoints support all typical features, including conditions. Stack access includes the ability to evaluate guest language expressions in halted contexts, object inspection, and setting of both local variables and object members. Protocols are implemented so that these services can be accessed by Chrome DevTools [15], the NetBeans IDE [28], and GraalVM's experimental REPL-style shell. These work uniformly for all Truffle languages, even when debugging polyglot programs where a stack may contain code in multiple languages.

### 7.2.2 Dynamic Profiling
GraalVM also supports the Chrome Inspector's Profiler Domain, as specified by the v8 version of the protocol (including CPU Profiling, Code Coverage, and Type Profiling) [14].

NodeProf [37] uses instrumentation to observe expression evaluations and to trace values for a variety of dynamic analyses, including most recently an alternate implementation of Jalangi [35] for locating performance problems in Node.js programs. NodeProf addresses several limitations in the standard code-rewriting implementation: NodeProf can analyze builtin and library code; NodeProf does not affect program semantics; NodeProf has no overhead until activated ("hot plugged") and none shortly after deactivation; and NodeProf can run significantly faster.

### 7.2.3 Language Implementation
Some language features that typically rely heavily on reflection have been addressed by instrumentation:

- A Ruby programmer may at any time call `set_trace_func`, which requires that a specified block of code be executed dynamically before each statement. This feature is notorious for disrupting optimization, but it runs fully optimized using Truffle Instrumentation as measured in Appendix B.
- Python's `settrace` command is implemented much like Ruby's `set_trace_func`.
- The R language includes an interactive shell that must be prepared at any time to enable *stepping* through specified methods, which is easily addressed using techniques similar to those used in the Debugging API.

### 7.2.4 GraalVM Implementation

Although originally conceived as a support layer specifically for programming tools, Truffle Instrumentation now supports a growing number GraalVM features, some originally unanticipated, for example:

- Constraints on program execution by time, statement count, or memory allocations can be implemented by installing an instrument to count the resource and possibly use built-in debugging support to terminate execution if needed.
- An http-based serviceability *agent* with access to built-in debugging/profiling services that incurs near zero overhead in fully optimized code until activated. Overhead returns to near zero when deactivated.

### 7.2.5 Outside clients

A growing number of instrumentation-supported tools have been developed outside Oracle Labs, including some unknown to the authors until published.

- Various run-time metrics were collected to assess the dynamic behavior of benchmarks for a cross-language comparison of the effectiveness of compilers [21].
- The Kómpos debugger addresses concurrency issues. It adds support for high-level breakpoints and stepping in the context of as asynchronous message sends, fork/join operations, and transactions. This allows developers to debug their programs at the same conceptual level used to implement their algorithms [22].
- An event profiling framework was applied to Truffle implementations of Python and Ruby, allowing cross-language comparisons of benchmark implementations [31].
- Truffle Instrumentation was used and extended for three implementations (increasing in generality) of dynamic method reloading [29].

## 8 Related work

Smalltalk [13] and Self [2], like Truffle, implemented debugging at the level of language abstractions. Being single-language systems, they simply inserted "halt" statements. The Truffle Instrumentation API generalizes to multiple languages, multiple kinds of tools, and multiple concurrently active non-interfering tools.

Self also pioneered full service debugging in dynamically optimized code with the addition of *deoptimization*. The compiler was extended to store just enough information to restore full access to execution state when needed [18]. Deoptimization was originally built into GraalVM for another purpose: supporting optimization by speculation [7]. Truffle Instrumentation minimizes overhead by managing speculation and deoptimization when inserting and removing instrumentation nodes.

Tool access to Java execution state is often implemented by rewriting bytecode, and early work in dynamic bytecode modification acknowledged its potential application to tools [6]. In the context of GraalVM, however, bytecode level support would not reflect guest language semantics. Truffle Instrumentation is part of GraalVM's *language implementation framework*, not the execution engine. Client APIs are expressed in terms of guest language source text and program elements, not bytecode, which distinguishes this approach from tooling for languages implemented on the JVM [26] or OMR [10]. Finally, unlike bytcode rewriting, multiple concurrent instrumentation clients do not interfere with one another.

The Maxine Inspector is a dedicated debugger and heap inspector for the metacircular Maxine Research VM [39]. Like Truffle Instrumentation, it is built using abstractions of the VM implementation, with which it shares a considerable amount of code. Unlike Truffle, the Inspector is designed only for debugging the VM itself.

There have been attempts to automate the delivery of debuggers and other tools in language-agnostic frameworks. Some impose important constraints on language implementation, for example the framework for debugging proposed by Wu, Gray, and Mernik [40]. Others lack the performance advantages of close runtime integration, for example debuggers generated by the Spoofax language workbench [20].

Interacting with execution events in managed runtime environments generally requires intrusive VM modifications, for example *meta-programming* infrastructure for Aspect Oriented Programming [26]. These typically target specific AOP approaches as narrowly as possible in order to minimize performance cost and complexity. A flexible *interposition*-based model for built-in VM support of AOP program was proposed [17], but only realized in a low-performance experiment to demonstrate generality [32]. Truffle Instrumentation instrumentation also uses flexible interposition, realized in the context of Truffle by dynamic injection of nodes into ASTs under interpretation and with careful attention to making instrumentation nodes fully optimizeable.

A Truffle experiment demonstrated the potential for low-overhead node-based interposition, prototyped by a simple in-language debugger for Ruby [34].

Neverlang supports an experimental implementation of "open programming language interpreters", allowing clients to "introspect, modify and extend the system's standard behaviour" [1]. Some stated goals are similar to Truffle's, for example, "linguistic tooling and instrumentation (e.g., debuggers)", but Neverlang is much more general. Truffle interpreters are not "open", and Truffle Instrumentation is designed to be orthogonal to language semantics in all but a few special cases for debugging. For example, Neverlang permits the dynamic addition of tree-rewriting rules, but tree-rewriting in Truffle Instrumentation is limited to highly optimized node wrapping/unwrapping for capturing AST execution events. Generalized tree rewriting also invites interference among multiple clients, which Truffle precludes by design.

## 9 Status and future work

Applications and extensions of the Truffle Instrumentation Framework continue to be developed, often with academic collaborators.

The Debugging API is being extended with a highly configurable way to unwind the execution stack and modify execution. This can be done at the granularity of method calls (stack "popping") or at the much finer granularity of designated AST nodes, for example expressions. Possible actions include re-entering or replacing its return value.

NodeProf's implementation of the Jalangi framework for dynamic analysis of Node.js programs [35], mentioned in Section 7, is being extended to other Truffle languages [37].

A Truffle implementation of the Language Server Protocol [24] is under development. This will ensure that a growing number of editing tools will fully support auto-completion and other language-sensitive features for all Truffle languages.

Finally, we plan to continue supporting and encouraging open-ended experimentation with tools that might otherwise require prohibitively difficult VM modification.

## 10   Conclusions

We have proposed strategies for managing the costs of ensuring that programmers "have it all" with respect to runtime-based tools: minimally disruptive, always available for every language and in every execution context. The Truffle Instrumentation Framework, an extension to GraalVM's Truffle Language Implementation Framework, is a proof of concept that this is possible.

Low overhead, flexible support for tools is now built into every Truffle language implementation, based on a framework of language-agnostic code that requires minimal per-language specialization. A growing collection of tools such as debuggers, profilers, and dynamic analysis frameworks make use of that support via Java APIs. Those APIs are expressed at the abstraction level of guest language structure and execution events, not VM implementation details or representations such as bytecode. Some of those tools are now part of GraalVM, and experimental tools are in development, both inside and outside Oracle Labs.

We have yet to find any reason why Truffle Instrumentation or services that use it should ever be disabled.

**Acknowledgements**   We are indebted to members of the Virtual Machine Research Group at Oracle Labs and the Institute of System Software at the Johannes Kepler University Linz for creating the language implementation technologies that make this work possible. Mario Wolczko and David Leibs of Oracle Labs provided foundational insights and encouragement throughout. Martin Entlicher, while a member of Oracle's NetBeans team, created the first IDE integration with the Truffle Debugging API and now, as a member of the Truffle team, provides ongoing development of the Debugging API and contributes to Truffle Instrumentation. Jaroslav Tulach facilitated our NetBeans collaboration and provided ongoing API guidance. David Piorkowsky added a valuable human factors touch to the Instrumentation API design during a 2014 internship. Tom Rodriguez provided helpful insight about JVMTI. Yuval Peduel, Stefan Marr, Sarah Mount, Laurence Tratt, and anonymous reviewers provided valuable feedback on drafts of this paper.

## A  Subscription maintenance

■ **Listing 11**  Subscription maintenance pseudocode

```
1  def rootCheck(AST, binding)
2          // could binding match any source locations in AST?
3
4  def addAST(AST)
5          for all bindings
6                  if rootCheck(AST, binding)
7                          for all nodes in AST
8                                  if match(node, binding)
9                                          probe = probe(node)
10                                         invalidate(probe)
11
12 def addBinding(binding)
13         for all ASTs
14                 if rootCheck(AST, binding)
15                         for all nodes in AST
16                                 if match(node, binding)
17                                         probe = getProbe(node)
18                                         if probe == null
19                                                 probe = probe(node)
20                                         invalidate(probe)
21
22 def removeBinding(binding)
23         for all ASTs
24                 if rootCheck(AST, binding)
25                         for all nodes in AST
26                                 if match(node, binding)
27                                         assert probed(node)
28                                         invalidate(probe)
29
30 def checkSubscriptions(probe)
31         if invalid(probe)
32                 remove all subscription nodes
33                 node = getNode(probe)
34                 for all bindings
35                         if match(node, binding)
36                                 addSubscription(probe, binding)
37                 if number of subscriptions > 0
38                         validate(probe)
39                 else
40                         remove(probe)
```

## B  Probe overhead

We evaluated performance characteristics of the Truffle Instrumentation Framework with two use cases in the Ruby programming language [9]: implementing set_trace_func, a standard feature of the language, and building a simple debugger that sets breakpoints.

Our experimental system is a Sun X4-2 server running two Intel Xeon E5-2660 Ivybridge CPUs with 8 cores and 16 hardware threads each at 2.20GHz, 256 GB of RAM, and running Oracle Linux Server 6.5. We used Graal 05845, based on OpenJDK 1.8.0_111 with JVMCI 0.23.

We compared TruffleRuby, the GraalVM implementation, against JRuby [25], an implementation using conventional JVM technology. Both implementations live in the same repository at commit 379e8.

We used a simple Mandelbrot program from the Computer Language Benchmarks Game.

In JRuby set_trace_func is disabled by default, so we first measured baseline performance with tracing disabled. We ran it for four minutes and took the mean average of the iteration times for the last two minutes to very generously allow for warmup. In TruffleRuby set_trace_func is always enabled, so tracing was already enabled in that first measurement.

Our first experiment used Ruby's set_trace_func feature, which installs a callback to be run each time the interpreter arrives at a new source line. With both implementations in turn, we ran four additional measurements:

- with set_trace_func enabled but unused;
- then with an empty callback installed;
- then with the call back replaced by one that increments a global variable; and finally
- with all callbacks removed.

Table 1 shows results with mean average time per iteration shown in seconds (so lower is better) and one standard deviation as an error. We can see that the baseline performance of TruffleRuby is an order of magnitude better than JRuby when tracing is disabled. TruffleRuby tracing is always enabled, so there is no impact on performance until a callback is installed. In JRuby, however, enabling tracing disables the just-in-time to bytecode compiler, which does not support tracing.

Installing an empty trace callback reduces performance in JRuby by another order of magnitude on top of the existing slowdown, but only slows TruffleRuby to half speed. A callback that increments a global variable slows both JRuby and TruffleRuby, but in this state TruffleRuby is still 100x faster than JRuby. When the callback is removed entirely TruffleRuby performance returns to the baseline level, but JRuby performance appears to suffer permanently.

An empty trace in TruffleRuby still has an impact because the Ruby logic for calling any block has some overhead, such as checking the mutable class of the block for the correct way to call it. If the trace instrument is modified to not call the trace block

■ **Table 1**  Performance times for set_trace_func, lower is better

|  | Disabled | Before | Empty | Increment | After |
|---|---|---|---|---|---|
| JRuby | 0.555 ±0.004 | 15.928 ±0.062 | 125.371 ±0.0 | 338.526 ±0.0 | 16.707 ±0.047 |
| TruffleRuby | 0.044 ±0.001 | 0.044 ±0.001 | 0.085 ±0.001 | 2.096 ±0.006 | 0.044 ±0.0 |

■ **Table 2**  Performance times Ruby debugging, lower is better

|  | Disabled | Before | Not-taken | Conditional | After |
|---|---|---|---|---|---|
| JRuby | 0.555 ±0.004 | 14.39 ±0.725 | 37.503 ±0.023 | 45.368 ±0.03 | 39.004 ±0.082 |
| TruffleRuby | 0.044 ±0.001 | 0.044 ±0.001 | 0.044 ±0.0 | 0.044 ±0.0 | 0.044 ±0.0 |

then the generated machine code for methods is exactly the same with the instrument installed or not.

In a second experiment, we modified the same Mandelbrot program to start a debugger which we used to insert breakpoints. In JRuby we used the *ruby-debug* module, a Java extension promoted as a faster alternative to historical Ruby debuggers. In TruffleRuby we implemented the same functionality using the Instrumentation Framework.

Similar to the previous experiment, we measured performance with debugging disabled, then with debugging enabled but no breakpoints installed. We then installed a breakpoint on a line on a branch that is not taken as the benchmark runs. Next we install a conditional breakpoint on a line that is taken, but for which the condition is never true. Finally, we removed all the breakpoints.

Similar to the previous experiment, results in Table 2 show that simply enabling debugging decimates JRuby performance. Installing a breakpoint that is never reached further reduces performance even though it is not actually in the path of execution. The extra work to test the breakpoint condition adds a further penalty. After all breakpoints are removed performance seems to remain permanently reduced. In TruffleRuby there is no performance impact to enabling debugging, as it is enabled by default. A breakpoint on a line never reached has no impact on performance, and the expression in the conditional breakpoint is inlined into the generated machine code for the method and is not measurable. When all breakpoints are removed, performance remains at the level as if a debugger had never been attached.

## About the authors

**Michael L. Van De Vanter** is a Researcher at Oracle Labs. His research interests include any kinds of tools that help programmers be more productive. Contact him at mlvdv@acm.org or via vandevanter.net/mlvdv.

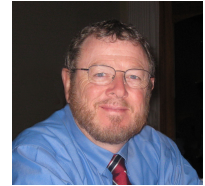

**Chris Seaton** is a Research Manager at Oracle Labs. Contact him at chris.seaton@oracle.com.

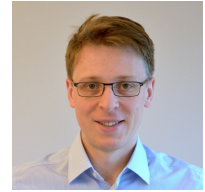

**Michael Haupt** is an Engineering Manager at eBay. He contributed to this work while at Oracle Labs and Oracle's Java Platform Group. Contact him at haupt@acm.org.

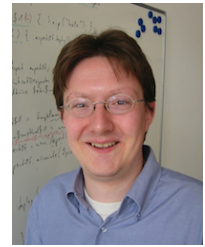

**Christian Humer** is member of the Oracle Labs VM Research Group. He is working on Graal and Truffle. Contact him at christian.humer@oracle.com.

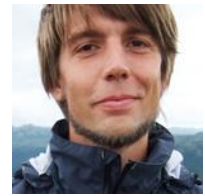

**Thomas Würthinger** is a Senior Research Director at Oracle Labs. Contact him at thomas.wuerthinger@oracle.com.

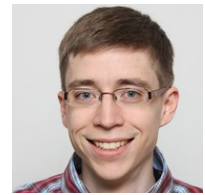